\documentclass[journal]{IEEEmytran}
\usepackage{amsfonts}
\usepackage{amssymb}

\newtheorem{theorem}{Theorem}[section]
\newtheorem{lemma}[theorem]{Lemma}
\newtheorem{proposition}[theorem]{Proposition}
\newtheorem{corollary}[theorem]{Corollary}
\newtheorem{EXAMPLE}{Example}[section]
\newenvironment{example}{\begin{EXAMPLE}\rm}{\rm\end{EXAMPLE}}
\newenvironment{algorithm}{%
        \setlength{\Algwidth}{\textwidth}%
        \addtolength{\Algwidth}{-2em}  
        \makebox[0ex]{}\hrulefill\makebox[0ex]{}\\[1.5ex]%
        \centering\begin{minipage}{\Algwidth}%
        \vspace{0.5ex}\normalsize}{\end{minipage}\vspace{2ex}%
        \par\makebox[0ex]{}\hrulefill\makebox[0ex]{}}
\newcommand{\graph}{{\mathcal{G}}}
\newcommand{\code}{{\mathcal{C}}}
\newcommand{\Code}{{\mathsf{C}}}
\newcommand{\BigCode}{{\mathbb{C}}}
\newcommand{\Real}{{\mathbb{R}}}
\newcommand{\erasure}{{\mathrm{?}}}
\newcommand{\encoder}{{\mathcal{E}}} 
\newcommand{\decoder}{{\mathcal{D}}}
\newcommand{\ngbr}{{\mathcal{N}}}
\newcommand{\GF}{{\mathrm{GF}}}
\newcommand{\Int}{{\mathrm{m}}}
\newcommand{\bldc}{{\mbox{\boldmath $c$}}}
\newcommand{\bldd}{{\mbox{\boldmath $d$}}}
\newcommand{\bldh}{{\mbox{\boldmath $h$}}}
\newcommand{\blds}{{\mbox{\boldmath $s$}}}
\newcommand{\bldt}{{\mbox{\boldmath $t$}}}
\newcommand{\bldv}{{\mbox{\boldmath $v$}}}
\newcommand{\bldw}{{\mbox{\boldmath $w$}}}
\newcommand{\bldx}{{\mbox{\boldmath $x$}}}
\newcommand{\bldy}{{\mbox{\boldmath $y$}}}
\newcommand{\bldz}{{\mbox{\boldmath $z$}}}
\newcommand{\bldeta}{{\mbox{\boldmath $\eta$}}}
\newcommand{\expected}{{\mathsf{E}}}
\newcommand{\variance}{{\mathsf{Var}}}
\newcommand{\Poly}{\textsc{Poly}}
\newcommand{\Eg}{{\expected_\graph}}
\newcommand{\EAg}{{\expected'_\graph}}
\newcommand{\EBg}{{\expected''_\graph}}
\newcommand{\VarAg}{{\variance'_\graph}}
\newcommand{\VarBg}{{\variance''_\graph}}
\newcommand{\bldone}{{\mathbf{1}}}
\newcommand{\half}{{\textstyle\frac{1}{2}}}
\newcommand{\qed}{\hspace*{\fill}%
    \vbox{\hrule\hbox{\vrule\squarebox{.667em}\vrule}\hrule}\smallskip}
    \def\squarebox#1{\hbox to #1{\hfill\vbox to #1{\vfill}}}

\newlength{\Initlabel}
\newlength{\Algwidth}

\title{Improved Nearly-MDS Expander Codes%
    \thanks{%
    The authors are with the Computer Science Department,
    Technion, Haifa 32000, Israel,
    e-mail: \{ronny, vitalys\}@cs.technion.ac.il.
    This work was supported by
    the Israel Science Foundation (Grant No.~746/04).
    Part of this work was presented at the
    \textit{2004 IEEE Int'l Symposium on Information Theory
    (ISIT'2004),}
    Chicago, Illinois (June 2004).}}
\author{Ron M. Roth \, and \, Vitaly Skachek}

\begin{document}
\maketitle

\begin{abstract}
A construction of expander codes is presented with
the following three properties:
(i)~the codes lie close to the Singleton bound,
(ii)~they can be encoded in time complexity that is linear in their
code length,
and
(iii)~they have a linear-time bounded-distance decoder.
By using a version of the decoder that corrects also erasures,
the codes can replace MDS outer codes in
concatenated constructions, thus resulting in 
linear-time encodable and decodable codes that approach
the Zyablov bound or the capacity of memoryless channels.
The presented construction improves on an earlier result by
Guruswami and Indyk in that
any rate and relative minimum distance that lies below the Singleton
bound is attainable for a significantly smaller alphabet size.

\end{abstract}

\section{Introduction}
\label{sec:introduction}

In this work, we consider a family of codes that are based
on expander graphs. The notion of graph codes was introduced by
Tanner in~\cite{Tanner1}.
Later,
the explicit constructions of Ramanujan expander graphs due to
Lubotsky, Philips, and Sarnak~\cite[Chapter~4]{DSV}, \cite{LPS}
and Margulis~\cite{Margulis}, were used by
Alon \emph{et al.} in~\cite{ABNNR} as building blocks
to obtain new polynomial-time constructions of asymptotically good codes
in the low-rate range
(by ``asymptotically good codes'' we mean codes whose rate
and relative minimum distance are both bounded away from zero).
Expander graphs were used then by Sipser and Spielman
in~\cite{SipSpielman} to present polynomial-time constructions
of asymptotically good codes that can be decoded
in time complexity which is linear in the code length.
By combining ideas from~\cite{ABNNR} and~\cite{SipSpielman},
Spielman provided in~\cite{Spielman} an asymptotically good construction
where both the decoding and encoding time complexities
were linear in the code length.

While the linear-time decoder of the Sipser-Spielman construction was
guaranteed to correct a number of errors that is a positive fraction of
the code length, that fraction was significantly smaller than what
one could attain by bounded-distance decoding---namely,
decoding up to half the minimum distance of the code.
The guaranteed fraction of linear-time correctable errors
was substantially improved by Z\'{e}mor in~\cite{Zemor01}.
In his analysis, Z\'{e}mor considered the special (yet abundant) case of
the Sipser-Spielman construction where
the underlying Ramanujan graph is bipartite, and presented
a linear-time iterative decoder where the correctable fraction
was $1/4$ of the relative minimum distance of the code.
An additional improvement by a factor of two, which brought
the (linear-time correctable) fraction to be essentially equal
to that of bounded-distance decoding,
was then achieved by the authors of this paper
in~\cite{Skachek-Roth-2003},
where the iterative decoder of Z\'{e}mor was enhanced
through a technique akin to generalized minimum distance (GMD)
decoding~\cite{ForneyMonograph}, \cite{Forney}.

In~\cite{GI3}, Guruswami and Indyk used
Z\'{e}mor's construction as a building block and combined it with
methods from~\cite{ABNNR}, \cite{AEL}, and~\cite{AlonLuby}
to suggest a code construction with the following three properties:
\begin{list}{}{\settowidth{\labelwidth}{(PX)}}
\item[(P1)]
The construction is nearly-MDS:
it yields
for every designed rate $R \in (0,1]$ and sufficiently small $\epsilon > 0$
an infinite family of codes of rate at least $R$
over an alphabet of size
\begin{equation}
\label{eq:GIalphabetsize}
2^{O\left((\log(1/\epsilon))/(R \epsilon^4) \right)} \; ,
\end{equation}
and the relative minimum distance of the codes is greater than
\[
1 - R - \epsilon \; .
\]
\item[(P2)]
The construction is linear-time encodable, and
the time complexity per symbol is $\Poly(1/\epsilon)$
(i.e., this complexity grows polynomially with $1/\epsilon$).
\item[(P3)]
The construction has a linear-time decoder
which is essentially a bounded-distance decoder:
the correctable number of errors is at least
a fraction $(1{-}R{-}\epsilon)/2$ of the code length.
The time complexity per symbol of
the decoder is also $\Poly(1/\epsilon)$.
\end{list}
In fact, the decoder described by Guruswami and Indyk in~\cite{GI3}
is more general in that it can handle
a combination of errors and erasures. Thus, by using their codes as
an outer code in a concatenated construction,
one obtains a linear-time encodable code that attains
the Zyablov bound~\cite[p.~1949]{Dumer},
with a linear-time bounded-distance decoder.
Alternatively, such a concatenated construction approaches
the capacity of any given memoryless channel:
if the inner code is taken to have the smallest decoding error exponent,
then the overall decoding error probability behaves like
Forney's error exponent~\cite{ForneyMonograph}, \cite{Forney}
(the time complexity of searching for the inner code, in turn, 
depends on $\epsilon$, yet not on the overall length of
the concatenated code).

Codes with similar attributes,
both with respect to the Zyablov bound and to the capacity of
memoryless channels, were presented also by Barg and Z\'{e}mor
in a sequence of papers~\cite{Zemor02},
\cite{Zemor-DIMACS}, \cite{Zemor03}
(yet in their constructions, only the decoding is guaranteed
to be linear-time).

In this work, we present a family of codes which improves
on the Guruswami-Indyk construction.
Specifically, our codes will satisfy properties (P1)--(P3),
except that the alphabet size in property~(P1) will now be only
\begin{equation}
\label{eq:alphabetsize}
2^{O\left((\log(1/\epsilon))/\epsilon^3 \right)} \; .
\end{equation}
The basic ingredients of our construction are similar to those
used in~\cite{GI3} (and also in~\cite{AEL} and~\cite{AlonLuby}),
yet their layout (in particular, the order of application of the 
various building blocks), 
and the choice of parameters will be different.
Our presentation will be split into two parts. We first describe
in Section~\ref{sec:firstconstruction} a construction that satisfies
only the two properties~(P1) and~(P3) over an alphabet of
size~(\ref{eq:alphabetsize}). 
These two properties will be proved
in Sections~\ref{sec:boundsonparameters} and~\ref{sec:decoding}.
We also show that the codes studied by
Barg and Z\'{e}mor in~\cite{Zemor02} and~\cite{Zemor03}
can be seen as concatenated codes, with our codes serving
as the outer codes.

The second part of our presentation consists of
Section~\ref{sec:lineartimeencoding}, where we modify
the construction of Section~\ref{sec:firstconstruction}
and use the resulting code as a building block
in a second construction, which satisfies property~(P2) as well.

\section{Construction of linear-time decodable codes}
\label{sec:firstconstruction}

Let $\graph = (V':V'',E)$ be 
a bipartite $\Delta$-regular undirected connected graph with
a vertex set $V = V' \cup V''$ such that $V' \cap V'' = \emptyset$,
and an edge set $E$ such that
every edge in $E$ has one endpoint in $V'$ and one endpoint in $V''$.
We denote the size of $V'$ by $n$
(clearly, $n$ is also the size of $V''$) and we will assume hereafter
without any practical loss of generality that $n > 1$.
For every vertex $u \in V$, we denote by $E(u)$ the set of
edges that are incident with $u$.
We assume an ordering on $V$, thereby inducing
an ordering on the edges of $E(u)$ for every $u \in V$.
For an alphabet $F$ and a word $\bldz = (z_e)_{e \in E}$ 
(whose entries are indexed by $E$) in $F^{|E|}$,
we denote by $(\bldz)_{\!\scriptscriptstyle E(u)}$
the sub-block of $\bldz$ that is indexed by $E(u)$.

Let $F$ be the field $\GF(q)$ and let $\code'$ and $\code''$ be
linear $[\Delta,r \Delta,\theta \Delta]$ and 
$[\Delta,R \Delta,\delta \Delta]$ codes over $F$, respectively.   
We define the code $\Code = (\graph, \code' : \code'')$
as the following linear code of length $|E|$ over $F$:
\begin{eqnarray*}
\lefteqn{
\Code = \Bigl\{ \bldc \in F^{|E|} \,:\,
\textrm{$(\bldc)_{\!\scriptscriptstyle E(u)} \in \code'$
for every $u \in V'$}} \makebox[15ex]{} \\
&&
\textrm{and $(\bldc)_{\!\scriptscriptstyle E(v)} \in \code''$
for every $v \in V''$} \Bigr\}
\end{eqnarray*}
($\Code$ is the primary code considered by Barg and Z\'{e}mor
in~\cite{Zemor02}).

Let $\Phi$ be the alphabet $F^{r \Delta}$.
Fix some linear one-to-one mapping
$\encoder : \Phi \rightarrow \code'$ over $F$, and let
the mapping $\psi_\encoder : \Code \rightarrow \Phi^n$ be given by
\begin{equation}
\label{eq:psi}
\psi_\encoder(\bldc) = 
\left(
\encoder^{-1} \left( (\bldc)_{\!\scriptscriptstyle E(u)} \right)
\right)_{u \in V'} \; , \quad \bldc \in \Code \; .
\end{equation}
That is, the entries of $\psi_\encoder(\bldc)$ are indexed
by $V'$, and the entry that is indexed by $u \in V'$
equals
$\encoder^{-1} \left( (\bldc)_{\!\scriptscriptstyle E(u)} \right)$.
We now define the code $(\Code)_\Phi$ of length $n$ over $\Phi$ by 
\[
(\Code)_\Phi =
\left\{ \psi_\encoder(\bldc) \,:\, \bldc \in \Code \right\} \; .
\]
Every codeword $\bldx = (\bldx_u)_{u \in V'}$ of $(\Code)_\Phi$
(with entries $\bldx_u$ in $\Phi$)
is associated with a unique codeword $\bldc \in \Code$ such that
\[
\encoder(\bldx_u) = (\bldc)_{\!\scriptscriptstyle E(u)} \; ,
\quad
u \in V' \; .
\]

Based on the definition of $(\Code)_\Phi$,
the code $\Code$ can be represented as
a concatenated code with an inner code $\code'$ over $F$ and
an outer code $(\Code)_\Phi$ over $\Phi$. 
It is possible, however, to use $(\Code)_\Phi$ as an outer code
with inner codes other than $\code'$.
Along these lines, the codes studied in~\cite{Zemor02}
and~\cite{Zemor03} can be 
represented as concatenated codes with $(\Code)_\Phi$ as an outer code,
whereas the inner codes are taken over a sub-field of $F$. 
 
\section{Bounds on the code parameters} 
\label{sec:boundsonparameters}

Let $\Code = (\graph, \code' :\code'')$, $\Phi$,
and $(\Code)_\Phi$ be as defined in Section~\ref{sec:firstconstruction}.
It was shown in~\cite{Zemor02} that the rate of $\Code$
is at least $r + R - 1$. From the fact that $\Code$ is
a concatenated code with an inner code $\code'$ and 
an outer code $(\Code)_\Phi$, it follows that
the rate of $(\Code)_\Phi$ is bounded from below by
\begin{equation}
\label{eq:rate_bnd}
\frac{r + R - 1}{r} =
1 - \frac{1}{r} + \frac{R}{r} \; .
\end{equation}
In particular, the rate approaches $R$ when $r \rightarrow 1$. 

We next turn to computing a lower bound on the relative minimum
distance of $(\Code)_\Phi$. By applying this lower bound,
we will then verify that $(\Code)_\Phi$ satisfies property~(P1).
Our analysis is based on that in~\cite{Zemor03},
and we obtain here an improvement over a bound that can be
inferred from~\cite{Zemor03}; we will need that improvement
to get the reduction of the alphabet size from~(\ref{eq:GIalphabetsize})
to~(\ref{eq:alphabetsize}). We first introduce several notations.

Denote by $A_\graph$ the adjacency matrix of $\graph$; namely,
$A_\graph$, is a $|V| \times |V|$ real symmetric matrix
whose rows and columns are indexed by the set $V$, and for every 
$u, v \in V$, the entry in $A_\graph$ that is indexed by $(u,v)$
is given by 
\[
(A_\graph)_{u,v} =
\left\{
\begin{array}{lcl} 
1 && \textrm{if $\{u, v\} \in E$} \\
0 && \textrm{otherwise} \\
\end{array}
\right. \; .
\]
It is known that $\Delta$ is the largest eigenvalue of $A_\graph$.
We denote by $\gamma_\graph$ the ratio 
between the second largest eigenvalue of $A_\graph$ and $\Delta$
(this ratio is less than $1$ when $\graph$ is connected
and is nonnegative when $n > 1$;
see~\cite[Propositions~1.1.2 and~1.1.4]{DSV}).

When $\graph$ is taken from a sequence of
Ramanujan expander graphs with constant degree $\Delta$,
such as the LPS graphs in~\cite{LPS}, we have
\[
\gamma_\graph \le \frac{2 \sqrt{\Delta{-}1}}{\Delta} \; .
\]

For a nonempty subset $S$ of the vertex set $V$ of $\graph$,
we will use the notation $\graph_S$ to stand for
the subgraph of $\graph$ that is induced by $S$:
the vertex set of $\graph_S$ is given by $S$,
and its edge set consists of all
the edges in $\graph$ that have each of their endpoints in $S$.
The degree of $u$ in $\graph_S$,
which is the number of adjacent vertices to $u$ in $\graph_S$,
will be denoted by $\deg_S(u)$.

\begin{theorem}
\label{thm:min_dist}
The relative minimum distance of the code $(\Code)_\Phi$ is 
bounded from below by 
\[
\frac{\delta - \gamma_\graph \sqrt{ \delta / \theta}}
{1 - \gamma_\graph} \; .
\]
In particular, this lower bound approaches $\delta$
when $\gamma_\graph \rightarrow 0$.
\end{theorem}

The proof of the theorem will make use of 
Proposition~\ref{prop:alonchung} below,
which is an improvement on Corollary~9.2.5 in
Alon and Spencer~\cite{AlonSpencer} for bipartite graphs,
and is also an improvement on Lemma~4 in Z\'{e}mor~\cite{Zemor01}.
We will need the following technical lemma for that proposition.
The proof of this lemma can be found in Appendix~\ref{app:A}.

Denote by $\ngbr(u)$ the set of vertices that are adjacent to vertex $u$ 
in $\graph$.

\begin{lemma}
\label{lemma:alonchung}
Let $\chi$ be a real function on the vertices of $\graph$
where the images of $\chi$ are restricted to the interval $[0,1]$.
Write 
\[
\sigma = \frac{1}{n} \sum_{u \in V'} \chi(u)
\qquad \textrm{and} \qquad 
\tau = \frac{1}{n} \sum_{v \in V''} \chi(v) \; .
\]
Then 
\begin{eqnarray*}
\frac{1}{\Delta n}
\sum_{u \in V'} \sum_{v \in \ngbr(u)} \chi(u) \chi(v)
& \le & 
\sigma \tau +
\gamma_\graph \sqrt{\sigma (1{-}\sigma) \tau (1{-}\tau)} \\
& \le &
(1{-}\gamma_\graph) \sigma \tau +
\gamma_\graph \sqrt{\sigma \tau} \; .
\end{eqnarray*}\vspace{0ex}
\end{lemma}
(Comparing to the results in~\cite{Zemor01}, Lemma 4 therein is stated for the special case 
where the images of $\chi$ are either $0$ or $1$. Our first inequality in Lemma~\ref{lemma:alonchung}
yields a bound which is always at least as tight as Lemma 4 in~\cite{Zemor01}.)

\begin{proposition}
\label{prop:alonchung}
Let $S \subseteq V'$ and $T \subseteq V''$ be subsets of sizes
$|S| = \sigma n$ and $|T| = \tau n$, respectively,
such that $\sigma + \tau > 0$.
Then the sum of the degrees in the graph $\graph_{S \cup T}$
is bounded from above by 
\[
\sum_{u \in S \cup T} \deg_{S \cup T}(u) \le
2 \left( (1{-}\gamma_\graph) \sigma \tau +  
\gamma_\graph \sqrt{\sigma \tau} \right) \Delta n \; .
\]
\end{proposition}

\emph{Proof:}
We select $\chi(u)$ in Lemma~\ref{lemma:alonchung} to be 
\[
\chi(u) =
\left\{
\begin{array}{lcl} 
1 && \textrm{if $u \in S \cup T$} \\
0 && \textrm{otherwise}
\end{array}
\right. \; .
\]
On the one hand, by Lemma~\ref{lemma:alonchung},
\[
\sum_{u \in V'} \sum_{v \in \ngbr(u)} \chi(u) \chi(v) \le  
\left( (1{-}\gamma_\graph) \sigma \tau +
\gamma_\graph \sqrt{\sigma \tau} \right) \Delta n \; .
\]
On the other hand,
\[
2
\sum_{u \in V'} \sum_{v \in \ngbr(u)} \chi(u) \chi(v) =
\sum_{u \in S \cup T} \deg_{S \cup T}(u) \; .
\]
These two equations yield the desired result.\qed

\emph{Proof of Theorem~\ref{thm:min_dist}:} 
First, it is easy to see that $(\Code)_\Phi$ is
a linear subspace over $F$ and, as such, it is
an Abelian subgroup of $\Phi^n$.
Thus, the minimum distance of $(\Code)_\Phi$ equals
the minimum weight (over $\Phi$) of any nonzero codeword of
$(\Code)_\Phi$. 

Pick any nonzero codeword $\bldx \in (\Code)_\Phi$,
and let $\bldc = (c_e)_{e \in E}$ be the unique codeword
in $\Code$ such that $\bldx = \psi_\encoder (\bldc)$.
Denote by $Y \subseteq E$ the support of $\bldc$ (over $F$), i.e.,
\[
Y = \{ e \in E \,:\, c_e \ne 0 \} \; .
\]
Let $S$ (respectively, $T$) be
the set of all vertices in $V'$ (respectively, $V''$)
that are endpoints of edges in $Y$.
In particular, $S$ is the support of the codeword $\bldx$.
Let $\sigma$ and $\tau$ denote
the ratios $|S|/n$ and $|T|/n$, respectively, and
consider the subgraph $\graph(Y) = (S:T,Y)$ of $\graph$. 
Since the minimum distance of $\code'$ is $\theta \Delta$,
the degree in $\graph(Y)$ of every vertex in $V'$ is at least
$\theta \Delta$. Therefore, the number of edges in $\graph(Y)$
satisfies
\[
|Y| \ge \theta \Delta \cdot \sigma n \; .
\]
Similarly, the degree in $\graph(Y)$ of every vertex in $V''$ is
at least $\delta \Delta$ and, thus,
\[
|Y| \ge \delta \Delta \cdot \tau n \; .
\]
Therefore,
\[
|Y| \ge \max \{ \theta \sigma, \delta \tau \} \cdot \Delta n \; . 
\]
On the other hand, $\graph(Y)$ is a subgraph of $\graph_{S \cup T}$;
hence, by Proposition~\ref{prop:alonchung},
\[
|Y| \le \frac{1}{2} \sum_{u \in S \cup T} \deg_{S \cup T}(u) \le
\left( (1{-}\gamma_\graph) \sigma \tau + \gamma_\graph 
\sqrt{\sigma \tau} \right) \Delta n \; .
\]
Combining the last two equations yields
\begin{equation}
\label{eq:Y}
\max \{ \theta \sigma, \delta \tau \} \le
(1{-}\gamma_\graph) \sigma \tau +
\gamma_\graph \sqrt{\sigma \tau} \; .
\end{equation}
We now distinguish between two cases.

\emph{Case 1: $\sigma/\tau \le \delta/\theta$.}
Here~(\ref{eq:Y}) becomes
\[
\delta \tau \le
(1{-}\gamma_\graph) \sigma \tau + \gamma_\graph \sqrt{\sigma \tau}
\]
and, so,
\begin{equation}
\label{eq:sigma_1}
\sigma \ge
\frac{\delta -
       \gamma_\graph \sqrt{\sigma / \tau}}{1 - \gamma_\graph} \ge 
\frac{\delta -
       \gamma_\graph \sqrt{\delta / \theta}}{1 - \gamma_\graph} \; .
\end{equation}

\emph{Case 2: $\sigma/\tau > \delta/\theta$.}
By exchanging between $\sigma$ and $\tau$
and between $\theta$ and $\delta$ in~(\ref{eq:sigma_1}), we get
\[
\tau \ge
\frac{\theta - \gamma_\graph \sqrt{\theta / \delta}}{1 - \gamma_\graph}
\; .
\]
Therefore, 
\[
\sigma >
\frac{\delta}{\theta} \cdot \tau \ge \frac{\delta}{\theta} \cdot
\frac{\theta - \gamma_\graph \sqrt{\theta/\delta}}{1 - \gamma_\graph} =
\frac{\delta - \gamma_\graph \sqrt{\delta/\theta}}{1 - \gamma_\graph}
\; .
\]

Either case yields the desired lower bound on the size,
$\sigma n$, of the support $S$ of $\bldx$.\qed

The next example demonstrates how the parameters of $(\Code)_\Phi$
can be tuned so that the improvement~(\ref{eq:alphabetsize}) of
property~(P1) holds.

\begin{example}
\label{ex:epsilon^3}
Fix $\theta = \epsilon$ for some small $\epsilon \in (0,1]$
(in which case $r > 1 - \epsilon$),
and then select $q$ and $\Delta$ so that
$q > \Delta \ge 4/\epsilon^3$.
For such parameters, we can take $\code'$ and $\code''$ to
be generalized Reed-Solomon (GRS) codes over $F$.
We also assume that $\graph$ is a Ramanujan bipartite graph,
in which case
\[
\gamma_\graph \le
\frac{2 \sqrt{\Delta{-}1}}{\Delta} < \epsilon^{3/2} \; .
\]
By~(\ref{eq:rate_bnd}),
the rate of $(\Code)_\Phi$ is bounded from below by
\[
1 - \frac{1}{1 - \epsilon} + \frac{R}{1 - \epsilon} >
R - \epsilon  \; ,
\]
and by Theorem~\ref{thm:min_dist}, the relative minimum distance
is at least
\begin{eqnarray*}
\frac{\delta -
         \gamma_\graph \sqrt{\delta / \theta}}{1 - \gamma_\graph}
& \ge &
\delta - \gamma_\graph \sqrt{\delta / \theta} >
\delta - \epsilon^{3/2} \cdot \frac{1}{\sqrt{\epsilon}} \\
& = &
\delta - \epsilon > 1{-}R{-}\epsilon \; . 
\end{eqnarray*}
Thus, the code $(\Code)_\Phi$ approaches the Singleton bound when
$\epsilon \rightarrow 0$.
In addition, if $q$ and $\Delta$ are selected to be (no larger than)
$O(1/\epsilon^3)$, then the alphabet $\Phi$ has size
\[
|\Phi| = q^{r \Delta} =
2^{O \left( (\log (1/\epsilon))/\epsilon^3 \right)} \; .
\] 
\qed
\end{example}
 From Example~\ref{ex:epsilon^3} we can state the following corollary. 
\begin{corollary}
\label{cor:alph-size}
For any designed rate $R \in (0, 1]$ and sufficiently small $\epsilon > 0$ there is an infinite family of codes 
$(\Code)_\Phi$ of rate at least $R$ and relative minimum distance greater than $1 - R - \epsilon$, 
over an alphabet of size as in~(\ref{eq:alphabetsize}). 
\end{corollary}

\section{Decoding algorithm}
\label{sec:decoding}

Let $\Code = (\graph,\code':\code'')$ be defined over $F = \GF(q)$ as
in Section~\ref{sec:firstconstruction}.
Figure~\ref{fig:decoder} presents
an adaptation of the iterative decoder
of Sipser and Spielman~\cite{SipSpielman} and Z\'{e}mor~\cite{Zemor01}
to the code $(\Code)_\Phi$, with the additional feature of handling
erasures (as well as errors over $\Phi$):
as we show in Theorem~\ref{thm:main} below, the algorithm corrects
any pattern of $t$ errors and $\rho$ erasures,
provided that $t + (\rho/2) < \beta n$, where
\[
\beta = \frac{(\delta/2) -
       \gamma_\graph \sqrt{\delta/\theta}}{1 - \gamma_\graph} \; .
\]
Note that $\beta$ equals approximately half
the lower bound in Theorem~\ref{thm:min_dist}. 
The value of $\nu$ in the algorithm,
which is specified in Theorem~\ref{thm:main} below,
grows logarithmically with $n$.

\begin{figure*}
\begin{algorithm}
\newcommand{\Item}[1]{\item[{\textbf{#1}\hfill}]}
\settowidth{\Initlabel}{\textbf{Initialize:}}
\begin{list}{}{\setlength{\labelwidth}{\Initlabel}%
               \setlength{\leftmargin}{\Initlabel}%
               \addtolength{\itemsep}{0ex}}
\Item{Input:}
Received word $\bldy = (\bldy_u)_{u \in V'}$
in $(\Phi \cup \{ \erasure \})^n$.
\Item{Initialize:}
For $u \in V'$ do:
$\quad \displaystyle (\bldz)_{\!\scriptscriptstyle E(u)} \leftarrow
\left\{
\begin{array}{lcl}
\encoder \left( \bldy_u \right)    && \textrm{if $\bldy_u \in \Phi$}  \\
\erasure \erasure \ldots \erasure  && \textrm{if $\bldy_u = \erasure$}
\end{array}
\right. \;$.
\Item{Iterate:}
For $i = 2, 3, \ldots, \nu$ do:
\begin{list}{}{\settowidth{\leftmargin}{(a)}%
               \addtolength{\leftmargin}{1.5em}%
               \settowidth{\labelwidth}{(a)}}
\item[(a)]
If $i$ is odd then
$U \equiv V'$ and $\decoder \equiv \decoder'$,
else $U \equiv V''$ and $\decoder \equiv \decoder''$.
\item[(b)]
For every $u \in U$ do:
$(\bldz)_{\!\scriptscriptstyle E(u)} \leftarrow
\decoder \left( (\bldz)_{\!\scriptscriptstyle E(u)} \right)$.
\end{list}
\Item{Output:}
$\psi_\encoder(\bldz)$ if $\bldz \in \Code$
(and declare `error' otherwise).
\end{list}
\end{algorithm}
\caption{Decoder for $(\Code)_\Phi$.}
\label{fig:decoder}
\end{figure*}

We use the notation ``$\erasure$'' to stand for an erasure.
The algorithm in Figure~\ref{fig:decoder} makes use of a word 
$\bldz = (z_e)_{e \in E}$ over $F \cup \{ \erasure \}$
that is initialized
according to the contents of the received word $\bldy$ as follows.
Each sub-block $(\bldz)_{\!\scriptscriptstyle E(u)}$ that corresponds
to a non-erased entry $\bldy_u$ of $\bldy$
is initialized to the codeword $\encoder (\bldy_u)$ of $\code'$.
The remaining sub-blocks 
$(\bldz)_{\!\scriptscriptstyle E(u)}$ are initialized as erased words
of length $\Delta$.  
Iterations $i = 3, 5, 7, \ldots$ use an error-correcting decoder
$\decoder' : F^\Delta \rightarrow \code'$ that 
recovers correctly any pattern of less than $\theta \Delta/2$
errors (over $F$),
and iterations $i = 2, 4, 6, \ldots$ use
a combined error-erasure decoder
$\decoder'' : (F \cup \{ \erasure \})^\Delta \rightarrow \code''$
that recovers correctly any pattern of $a$ errors
and $b$ erasures, provided that $2a + b < \delta \Delta$
($b$ will be positive only when $i = 2$).

\begin{theorem} 
\label{thm:main}
Suppose that
\begin{equation}
\label{eq:exponentbase}
\sqrt{\theta \delta} > 2 \gamma_\graph > 0 \; , 
\end{equation}
and fix $\sigma$ to be a positive real number such that 
\begin{equation}
\label{eq:main}
\sigma < \beta = \frac{(\delta/2) -
      \gamma_\graph \sqrt{\delta/\theta}}{1 - \gamma_\graph} \; . 
\end{equation}
If
\[
\nu = 2 \Bigg \lfloor \log \left(
\frac{\beta \sqrt{\sigma n} - \sigma}{\beta - \sigma}
\right) \Bigg \rfloor  + 3
\]
then the decoder in Figure~\ref{fig:decoder} recovers correctly any
pattern of $t$ errors (over $\Phi$)
and $\rho$ erasures, provided that
\begin{equation}
\label{eq:maincondition}
t + \frac{\rho}{2} \le \sigma n \; .
\end{equation}\vspace{0ex}
\end{theorem}

The proof of the theorem makes use of the following lemma.

\begin{lemma}
\label{lemma:sqrt_st}
Let $\chi$, $\sigma$, and $\tau$ be as in Lemma~\ref{lemma:alonchung},
and suppose that
the restriction of $\chi$ to $V''$ is not identically zero
and that $\gamma_\graph > 0$.
Let $\delta$ be a real number for which the following condition 
is satisfied for every $v \in V''$:
\[
\chi(v) > 0 \; \Longrightarrow \;
\sum_{u \in \ngbr(v)} \chi(u) \ge \frac{\delta \Delta}{2}.  
\]
Then
\[
\sqrt{\frac{\sigma}{\tau}} \ge
\frac{ (\delta/2) - (1{-}\gamma_\graph) \sigma}{\gamma_\graph} \; .
\]\vspace{0ex}
\end{lemma}

The proof of Lemma~\ref{lemma:sqrt_st} can be found in
Appendix~\ref{app:A}.
This lemma implies an upper bound on $\tau$, in terms of $\sigma$;
it can be verified that this bound is always at least as tight
as Lemma~5 in~\cite{Zemor01}.  

\emph{Proof of Theorem~\ref{thm:main}:} 
For $i \ge 2$, let $U_i$ be the value of the set $U$ at the end of
iteration $i$ in Figure~\ref{fig:decoder}, and let $S_i$ be
the set of all vertices $u \in U_i$ such that 
$(\bldz)_{E(u)}$ is in error at the end of that iteration.
Let $\chi_1 : (V' \cup V'') \rightarrow \{0, \half, 1 \}$
be the function  
\[
\chi_1(u) =
\left\{
\begin{array}{lcl} 
1     && \textrm{if $u \in V'$ and $\bldy_u$ is in error}   \\
\half && \textrm{if $u \in V'$ and $\bldy_u$ is an erasure} \\
0     && \textrm{otherwise}
\end{array}
\right. \; ,
\]
and, for $i \ge 2$ define the function
$\chi_i : (V' \cup V'') \rightarrow \{ 0, \half, 1 \}$ recursively by 
\[
\chi_i(u) =
\left\{
\begin{array}{lcl} 
1             && \textrm{if $u \in S_i$} \\
0             && \textrm{if $u \in U_i \setminus S_i$} \\
\chi_{i-1}(u) && \textrm{if $u \in U_{i-1}$}
\end{array}
\right. \; ,
\]
where $U_1 = V'$. 

Denote 
\[
\sigma_i = \frac{1}{n} \sum_{u \in U_i} \chi_i(u) \; .
\]
Obviously, $\sigma_1 n = t + (\rho/2)$ and, so,
by~(\ref{eq:maincondition}) we have $\sigma_1 \le \sigma$.

Let $\ell$ be the smallest positive integer (possibly $\infty$)
such that $\sigma_\ell = 0$.
Since both $\decoder'$ and $\decoder''$ are bounded-distance
decoders, a vertex $v \in U_i$ can belong to $S_i$ for even $i \ge 2$,
only if the sum $\sum_{u \in \ngbr(v)} \chi_i(u)$
(which equals the sum $\sum_{u \in \ngbr(v)} \chi_{i-1}(u)$)
is at least $\delta \Delta/2$. Similarly,
a vertex $v \in U_i$ belongs to $S_i$ for odd $i > 1$, only if
$\sum_{u \in \ngbr(v)} \chi_i(u) \ge \theta \Delta/2$.
It follows that the function $\chi_i$ satisfies the conditions of
Lemma~\ref{lemma:sqrt_st}
(with $\theta$ taken instead of $\delta$ for odd $i$) and, so,
\begin{equation}
\label{eq:sigma_ratio}
\sqrt{\frac{\sigma_{i-1}}{\sigma_i}} \ge
\left\{
{\renewcommand{\arraystretch}{2.0}
\begin{array}{lcl} 
\displaystyle \frac{\delta}{2 \gamma_\graph} -
\frac{1{-}\gamma_\graph}{\gamma_\graph} \sigma_{i-1} 
&& \textrm{for even $0 < i < \ell$} \\
\displaystyle \frac{\theta}{2 \gamma_\graph} -
\frac{1{-}\gamma_\graph}{\gamma_\graph} \sigma_{i-1} 
&& \textrm{for odd $1 < i < \ell$}
\end{array}
}
\right. \; .
\end{equation}
Using the condition $\sigma_1 \le \sigma < \beta$,
it can be verified by induction on $i \ge 2$ that
\begin{equation}
\frac{\sigma_{i-1}}{\sigma_i}  \ge
\left\{
\begin{array}{lcl}
\delta/\theta && \textrm{for even $0 < i < \ell$} \\
\theta/\delta && \textrm{for odd  $ 1 < i < \ell$} 
\end{array}
\right. \; .
\label{eq:sigmas-ratio}
\end{equation}
Hence, for every $i > 2$,
\[
\frac{\sigma_{i-2}}{\sigma_i} =
\frac{\sigma_{i-2}}{\sigma_{i-1}} \cdot
\frac{\sigma_{i-1}}{\sigma_{i}} \ge
\frac{\delta}{\theta} \cdot \frac{\theta}{\delta} = 1 \; ;
\]
in particular, $\sigma_i \le \sigma$ for odd $i$
and $\sigma_i \le \sigma_2$ for even $i$.
Incorporating these inequalities into~(\ref{eq:sigma_ratio}) yields
\begin{equation}
\label{eq:ratio_sigma_2even}
\frac{1}{\sqrt{\sigma_i}} \ge
\frac{\delta}{2 \gamma_\graph \sqrt{\sigma_{i-1}}} -
\frac{1{-}\gamma_\graph}{\gamma_\graph} 
\sqrt{\sigma}
\qquad
\textrm{for even $0 < i < \ell$} \; \phantom{.}
\end{equation}
and
\begin{equation}
\label{eq:ratio_sigma_2odd}
\frac{1}{\sqrt{\sigma_i}} \ge
\frac{\theta}{2 \gamma_\graph \sqrt{\sigma_{i-1}}} -
\frac{1{-}\gamma_\graph}{\gamma_\graph}  
\sqrt{\sigma_2}
\qquad
\textrm{for odd $1 < i < \ell$}  \; .
\end{equation}

By combining~(\ref{eq:ratio_sigma_2even})
and~(\ref{eq:ratio_sigma_2odd}) we get that for even $i > 0$,
\begin{eqnarray*}
\lefteqn{
\frac{2 \gamma_\graph}{\theta \sqrt{\sigma_{i+1}}} +
\frac{2 (1{-}\gamma_\graph)}{\theta} \sqrt{\sigma_2}
\; \ge \;
\frac{1}{\sqrt{\sigma_i}} } \makebox[15ex]{} \\
& \ge &
\frac{\delta}{2 \gamma_\graph \sqrt{\sigma_{i-1}}} -
\frac{1{-}\gamma_\graph}{\gamma_\graph} \sqrt{\sigma} 
\; ,
\end{eqnarray*}
or
\begin{eqnarray}
\frac{1}{\sqrt{\sigma_{i+1}}}
& \ge & 
\frac{\theta \delta}{4 \gamma_\graph^2 \sqrt{\sigma_{i-1}}} - 
\frac{1{-}\gamma_\graph}{\gamma_\graph}
\left( \frac{\theta \sqrt{\sigma}}{2 \gamma_\graph} +
\sqrt{\sigma_2} \right) \nonumber \\
& \ge &
\frac{\theta \delta}{4 \gamma_\graph^2 \sqrt{\sigma_{i-1}}} - 
\frac{1{-}\gamma_\graph}{\gamma_\graph}
\left( \frac{\theta }{2 \gamma_\graph} +
\sqrt{\frac{\theta}{\delta}} \right) \sqrt{\sigma} \nonumber \\
& = &
\frac{\theta \delta}{4 \gamma_\graph^2}
\left( \frac{1}{\sqrt{\sigma_{i-1}}} - \frac{\sqrt{\sigma}}{\beta}
\right) + \frac{\sqrt{\sigma}}{\beta} \; ,
\label{eq:sigma-i+1}
\end{eqnarray}
where the second inequality follows from
$\sigma_2 \le \sigma \cdot \theta / \delta$
(see~(\ref{eq:sigmas-ratio})),
and the (last) equality follows from the next chain of equalities: 
\begin{eqnarray*}
\lefteqn{
\frac{1{-}\gamma_\graph}{\gamma_\graph}
\left( \frac{\theta}{2\gamma_\graph}
+ \sqrt{\frac{\theta}{\delta}} \right) \sqrt{\sigma}}
\makebox[10ex]{} \\
& = & 
\frac{1{-}\gamma_\graph}{2 \gamma_\graph^2} \left( 2\gamma_\graph  
+ \sqrt{\theta\delta} \right) \sqrt{\frac{\sigma \theta}{\delta}} \\
& = &
- \frac{1{-}\gamma_\graph}{2 \gamma_\graph^2} \cdot
\frac{ 4 \gamma_\graph^2   
- \theta \delta}{ \sqrt{\theta\delta} - 2\gamma_\graph}
\sqrt{\frac{\sigma \theta}{\delta}} \\
& = & - \left( 1- \frac{\theta \delta}{4 \gamma_\graph^2} \right) 
\frac{(1{-}\gamma_\graph) \sqrt{\sigma} }{ (\delta/2) - \gamma_\graph
\sqrt{\delta/\theta} } \\
& = &
- \left( 1- \frac{\theta \delta}{4 \gamma_\graph^2} \right) 
\frac{\sqrt{\sigma} }{\beta } \; . 
\end{eqnarray*}

Consider the following first-order linear recurring 
sequence $(\Lambda_j)_{j \ge 0}$ that satisfies 
\begin{eqnarray*}
\Lambda_{j+1} & = &
\displaystyle \frac{\theta \delta}{4 \gamma_\graph^2}
\left( \Lambda_j - \frac{\sqrt{\sigma}}{\beta} \right) 
+ \frac{\sqrt{\sigma}}{\beta} \; , \quad j \ge 0 \; , 
\end{eqnarray*}
where $\Lambda_{0} = 1 / \sqrt{\sigma}$. From~(\ref{eq:sigma-i+1}) we have
$1/\sqrt{\sigma_{i+1}} \ge \Lambda_{i/2}$ for even $i \ge 0$.
By solving the recurrence for 
$( \Lambda_j )$, we obtain 
\begin{eqnarray}
\frac{1}{\sqrt{\sigma_{i+1}}} \; \ge \; \Lambda_{i/2} =
\left( \left( \frac{\theta \delta}{4 \gamma_\graph^2} \right)^{i/2}
\left( 1 - \frac{\sigma}{\beta} \right)
+ \frac{\sigma}{\beta} \right) \frac{1}{\sqrt{\sigma}} \; . 
\label{eq:recurrence}
\end{eqnarray}

   From the condition~(\ref{eq:exponentbase}) we thus get
that $\sigma_{i+1}$ decreases exponentially with (even) $i$.
A sufficient condition for ending the decoding 
correctly after $\nu$ iterations is having $\sigma_\nu < 1/n$, or
\[
\frac{1}{\sqrt{\sigma_\nu}} > \sqrt{n} \; .  
\]
We require therefore that $\nu$ be such that 
\[
\frac{1}{\sqrt{\sigma_\nu}} \ge
\left( \left( \frac{\theta \delta}{4 \gamma_\graph^2}
\right)^{(\nu -1)/2} \left( 1 - \frac{\sigma}{\beta} \right)
+ \frac{\sigma}{\beta} \right) \frac{1}{\sqrt{\sigma}} > \sqrt{n} \; .
\]
The latter inequality can be rewritten as 
\[
\left( \frac{\theta \delta}{4 \gamma_\graph^2} \right)^{(\nu -1)/2} > 
\frac{\sqrt{n \sigma} - (\sigma / \beta)}{1 - (\sigma / \beta)} = 
\frac{\beta \sqrt{n \sigma} - \sigma}{\beta - \sigma} \; ,
\]
thus yielding 
\[
\nu > 2 \log\left( \frac{\beta \sqrt{n \sigma} - 
\sigma}{\beta - \sigma} \right) + 1 \; , 
\]
where the base of the logarithm equals
$(\theta \delta)/(4 \gamma_\graph^2)$. 
In summary, the decoding will end with the correct codeword after
\[
\nu = 2 \Bigg\lfloor \log \left( \frac{\beta \sqrt{n \sigma} - 
\sigma}{\beta - \sigma} \right) \Bigg\rfloor  + 3 
\] 
iterations (where the base of the logarithm again equals
$(\theta \delta)/(4 \gamma_\graph^2)$.)
\qed

In Lemma~\ref{lemma:complexity}, which appears in Appendix~\ref{app:B},
it is shown that the number of actual applications of the decoders $\decoder'$
and $\decoder''$ in the algorithm in Figure~\ref{fig:decoder} 
can be bounded from above by $\omega \cdot n$, where 
\[
\omega =  
2 \cdot 
\left\lceil \frac{\log\left( \frac{\displaystyle
\Delta \beta \sqrt{\sigma}}{ \displaystyle {\beta} - \sigma} \right)}
{\log \left( \displaystyle \frac{\theta \delta}{4 \gamma_\graph^2}
\right) } \right\rceil 
+ \frac{\displaystyle 1 + \frac{\theta}{\delta} } 
{\displaystyle 1 -
\left( \frac{4 \gamma_\graph^2}{\theta \delta} \right)^2 }  \; .
\]
Thus, if $\theta$ and $\delta$ are fixed and the ratio $\sigma/\beta$ is bounded away from $1$
and $\graph$ is a Ramanujan graph, then the value of $\omega$ is bounded from above by an 
absolute constant (independent of~$\Delta$). 

The algorithm in Figure~\ref{fig:decoder} allows us to use GMD decoding 
in cases where $(\Code)_\Phi$ is used as an outer code in
a concatenated code. In such a concatenated code,
the size of the inner code is $|\Phi|$ and, thus, it does not grow
with the length $n$ of $(\Code)_\Phi$.
A GMD decoder will apply the algorithm in Figure~\ref{fig:decoder}
a number of times that is proportional to the minimum distance of
the inner code. Thus, if the inner code has rate that is bounded
away from zero, then the GMD decoder will have time complexity
that grows linearly with the overall code length.
Furthermore, if $\code'$, $\code''$, and the inner code are codes
that have a polynomial-time bounded-distance decoder---e.g.,
if they are GRS codes---then the multiplying constant in
the linear expression of the time complexity
(when measured in operations in $F$) is $\Poly(\Delta)$.
For the choice of parameters in Example~\ref{ex:epsilon^3},
this constant is $\Poly(1/\epsilon)$
and, since $F$ is chosen in that example to have size $O(1/\epsilon^3)$,
each operation in $F$ can in turn be implemented by
$\Poly(\log(1/\epsilon))$ bit operations.
(We remark that in all our complexity estimates, we assume that
the graph $\graph$ is ``hard-wired'' so that we can ignore
the complexity of figuring out the set of incident edges of
a given vertex in $\graph$. Along these lines, we assume
that each access to an entry takes constant time,
even though the length of the index of that entry
may grow logarithmically with the code length.
See the discussion in~\cite[Section~II]{SipSpielman}.)

When the inner code is taken as $\code'$,
the concatenation results in the code $\Code = (\graph,\code':\code'')$ 
(of length  $\Delta n$) over $F$, and
the (linear-time) correctable fraction of errors is then
the product $\theta \cdot \sigma$,
for any positive real $\sigma$ that satisfies~(\ref{eq:main}).
A special case of this result, for $F = \GF(2)$ and $\code' = \code''$, 
was presented in our earlier work~\cite{Skachek-Roth-2003},
yet the analysis therein was different.
A linear-time decoder for $\Code$ was also presented
by Barg and Z\'{e}mor in~\cite{Zemor03},
except that their decoder requires finding a codeword that minimizes
some weighted distance function, and we are unaware of a method that
performs this task in time complexity that is
$\Poly(\Delta)$---even when $\code'$ and $\code''$
have a polynomial-time bounded-distance decoder.

\section{Construction which is also linear-time encodable}
\label{sec:lineartimeencoding}

In this section, we use the construction $(\Code)_\Phi$
of Section~\ref{sec:firstconstruction}
as a building block in obtaining a second construction,
which satisfies all properties (P1)--(P3) over an alphabet
whose size is given by~(\ref{eq:alphabetsize}). 

\subsection{Outline of the construction}
\label{sec:outline}

Let $\Code = (\graph,\code':\code'')$ be defined over $F = \GF(q)$ as
in Section~\ref{sec:firstconstruction}.
The first simple observation that provides the intuition behind
the upcoming construction is that the encoding of $\Code$,
and hence of $(\Code)_\Phi$, can be easily implemented in linear time
if the code $\code'$ has rate $r = 1$, in which case $\Phi = F^\Delta$.
The definition of $\Code$ then reduces to
\[
\Code = \left\{ \bldc \in F^{|E|} \,:\,
\textrm{$(\bldc)_{\!\scriptscriptstyle E(v)} \in \code''$
for every $v \in V''$} \right\} \; .
\]
We can implement an encoder of $\Code$ as follows.
Let $\encoder'' : F^{R \Delta} \rightarrow \code''$ be some
one-to-one encoding mapping of $\code''$.
Given an information word $\bldeta$ in $F^{R \Delta n}$,
it is first recast into a word of length $n$ over $F^{R \Delta}$
by sub-dividing it into sub-blocks $\bldeta_v \in F^{R \Delta}$
that are indexed by $v \in V''$;
then a codeword $\bldc \in \Code$ is computed by
\[
(\bldc)_{\!\scriptscriptstyle E(v)} = \encoder''(\bldeta_v) \; ,
\quad
v \in V'' \; .
\]
By selecting $\encoder$ in~(\ref{eq:psi}) as the identity mapping,
we get that the respective codeword
$\bldx = (\bldx_u)_{u \in V'} = \psi_\encoder(\bldc)$
in $(\Code)_\Phi$ is
\[
\bldx_u = (\bldc)_{\!\scriptscriptstyle E(u)} \; ,
\quad
u \in V' \; .
\]
Thus, each of the $\Delta$ entries (over $F$) of
the sub-block $\bldx_u$ can be associated with
a vertex $v \in \ngbr(u)$, and the value assigned to
that entry is equal to one of the entries in $\encoder''(\bldeta_v)$.

While having $\code' = \Phi \; (= F^\Delta)$ allows easy encoding,
the minimum distance of the resulting code $(\Code)_\Phi$ is
obviously poor.
To resolve this problem, we insert into the construction another linear
$[\Delta, r_0 \Delta, \theta_0 \Delta]$ code $\code_0$ over $F$.
Let $H_0$ be some $((1{-}r_0) \Delta) \times \Delta$
parity-check matrix of $\code_0$ and for
a vector $\bldh \in F^{(1-r_0)\Delta}$, denote by
$\code_0(\bldh)$ the following coset of $\code_0$ within $\Phi$:
\[
\code_0(\bldh) = 
\left\{ \bldv \in \Phi \,:\, H_0 \bldv = \bldh \right\} \; .
\]
Fix now a list of vectors $\blds = (\bldh_u)_{u \in V'}$
where $\bldh_u \in F^{(1-r_0)\Delta}$,
and define the subset $\Code(\blds)$ of $\Code$ by
\[
\Code(\blds) =
\left\{ \bldc \in \Code \,:\,
\textrm{$(\bldc)_{\!\scriptscriptstyle E(u)} \in \code_0(\bldh_u)$
for every $u \in V'$} \right\} \; ;
\]
accordingly, define
the subset $(\Code(\blds))_\Phi$ of $(\Code)_\Phi$ by 
\[
(\Code(\blds))_\Phi =
\Bigl\{
\psi_\encoder(\bldc) = 
\left( (\bldc)_{\!\scriptscriptstyle E(u)} \right)_{u \in V'} \,:\,
\bldc \in \Code(\blds) \Bigr\}
\; .
\]
Now, if $\blds$ is all-zero, then $\Code(\blds)$ coincides
with the code $\Code(\mathbf{0}) = (\graph,\code_0:\code'')$;
otherwise, $\Code(\blds)$ is either empty or is
a coset of $\Code(\mathbf{0})$, where $\Code(\mathbf{0})$ is regarded
as a linear subspace of $\Code$ over $F$. From
this observation we conclude that the lower bound
in Theorem~\ref{thm:min_dist} applies
to any nonempty subset $(\Code(\blds))_\Phi$, except that
we need to replace $\theta$ by $\theta_0$.

In addition, a simple modification in the algorithm in
Figure~\ref{fig:decoder} adapts it to decode
$(\Code(\blds))_\Phi$ so that Theorem~\ref{thm:main} holds
(again under the change $\theta \leftrightarrow \theta_0$):
during odd iterations $i$, we apply to
each sub-block $(\bldz)_{\!\scriptscriptstyle E(u)}$
a bounded-distance decoder of $\code_0(\bldh_u)$,
instead of the decoder $\decoder'$.

Therefore, our strategy in designing the linear-time encodable
codes will be as follows.
The raw data will first be encoded into a codeword
$\bldc$ of $\Code$ (where $\code' = \Phi$).
Then we compute the $n$ vectors
\[
\bldh_u = H_0 \cdot (\bldc)_{\!\scriptscriptstyle E(u)} \; ,
\quad
u \in V' \; ,
\]
and produce the list $\blds = (\bldh_u)_{u \in V'}$;
clearly, $\bldc$ belongs to $\Code(\blds)$.
The list $\blds$ will then undergo additional encoding stages,
and the result will be merged with $\psi_\encoder(\bldc)$
to produce the final codeword.
The parameters of $\code_0$, which determine the size of $\blds$,
will be chosen so that the overhead due to $\blds$ will be negligible.

During decoding, $\blds$ will be recovered first, and then 
we will apply the aforementioned adaptation to $(\Code(\blds))_\Phi$
of the decoder in Figure~\ref{fig:decoder},
to reconstruct the information word $\bldeta$.

\subsection{Details of the construction}

We now describe the construction in more detail.
We let $F$ be the field $\GF(q)$ and $\Delta_1$ and $\Delta_2$ be
positive integers.
The construction makes use of two bipartite regular graphs,
\[
\graph_1 = (V':V'',E_1)
\qquad
\textrm{and}
\qquad
\graph_2 = (V':V'',E_2) \; ,
\]
of degrees $\Delta_1$ and $\Delta_2$, respectively.
Both graphs have the same number of vertices; in fact, we are
making a stronger assumption whereby both graphs are defined over
the same set of vertices. We denote by $n$
the size of $V'$ (or $V''$) and by $\Phi_1$ and $\Phi_2$
the alphabets $F^{\Delta_1}$ and $F^{\Delta_2}$, respectively.
The notations $E_1(u)$ and $E_2(u)$ will stand for
the sets of edges that are incident with a vertex $u$
in $\graph_1$ and $\graph_2$, respectively.

We also assume that we have at our disposal the following four codes:
\begin{itemize}
\item
a linear $[\Delta_1, r_0 \Delta_1, \theta_0 \Delta_1]$ code
$\code_0$ over $F$;
\item
a linear $[\Delta_1, R_1 \Delta_1, \delta_1 \Delta_1]$ code
$\code_1$ over $F$;
\item
a linear $[\Delta_2, R_2 \Delta_2, \delta_2 \Delta_2]$ code
$\code_2$ over $F$;
\item
a code $\code_\Int$ of length $n$ and rate $r_\Int$ over
the alphabet $\Phi_\Int = F^{R_2 \Delta_2}$.
\end{itemize}
The rates of these codes need to satisfy the relation
\[
(1{-}r_0) \Delta_1 = r_\Int R_2 \Delta_2 \; ,
\]
and the code $\code_\Int$ is assumed to have the following properties:
\begin{enumerate}
\item
Its rate is bounded away from zero: there is
a universal positive constant $\kappa$ such that $r_\Int \ge \kappa$.
\item
$\code_\Int$ is linear-time encodable, and the encoding time
per symbol is $\Poly(\log |\Phi_\Int|)$.
\item
$\code_\Int$ has a decoder that recovers in linear-time any pattern of
up to $\mu n$ errors (over the alphabet $\Phi_\Int$),
where $\mu$ is a universal positive constant.
The time complexity per symbol of
the decoder is $\Poly(\log |\Phi_\Int|)$.
\end{enumerate}
(By a universal constant we mean a value that does not depend on
any other parameter, not even on the size of $\Phi_\Int$.)
For example, we can select as $\code_\Int$
the code of Spielman in~\cite{Spielman}, in which case
$\kappa$ can be taken as $1/4$.

Based on these ingredients, we introduce the codes
\[
\Code_1 = (\graph_1,\Phi_1:\code_1)
\qquad
\textrm{and}
\qquad
\Code_2 = (\graph_2,\Phi_2:\code_2)
\]
over $F$.
The code $\Code_1$ will play the role of the code $\Code$
as outlined in Section~\ref{sec:outline}, whereas
the codes $\code_\Int$ and $\Code_2$ will be utilized for
the encoding of the list $\blds$ that was described there.

The overall construction, which we denote by $\BigCode$,
is now defined as the set of all words of length $n$ over the alphabet
\[
\Phi = \Phi_1 \times \Phi_2
\]
that are obtained by applying
the encoding algorithm in Figure~\ref{fig:encoder}
to information words $\bldeta$ of length $n$ over $F^{R_1 \Delta_1}$.
A schematic diagram of the algorithm is shown
in Figure~\ref{fig:schematicdiagram}.
(In this algorithm, we use a notational convention whereby
entries of information words $\bldeta$ are indexed by $V''$,
and so are codewords of $\code_\Int$.)

\begin{figure*}
\begin{algorithm}
\settowidth{\Initlabel}{\textbf{Output:}}
\newcommand{\Item}[1]{\item[{\makebox[\Initlabel][c]{\textbf{#1}}}]}
\begin{list}{}{\setlength{\labelwidth}{\Initlabel}%
               \setlength{\leftmargin}{\Initlabel}%
               \addtolength{\itemsep}{1ex}}
\Item{Input:}
Information word $\bldeta = (\bldeta_v)_{v \in V''}$
of length $n$ over $F^{R_1 \Delta_1}$.
\Item{(E1)}
Using an encoder $\encoder_1 : F^{R_1 \Delta_1} \rightarrow \code_1$,
map $\bldeta$ into a codeword $\bldc$ of $\Code_1$ by
\[
(\bldc)_{\!\scriptscriptstyle E_1(v)} \leftarrow \encoder_1(\bldeta_v)
\; ,
\quad
v \in V'' \; .
\]
\Item{(E2)}
Fix some $((1{-}r_0) \Delta_1) \times \Delta_1$
parity-check matrix $H_0$ of $\code_0$ over $F$,
and compute the $n$ vectors
\[
\bldh_u \leftarrow H_0 \cdot (\bldc)_{\!\scriptscriptstyle E_1(u)} \; ,
\quad
u \in V' \; ,
\]
to produce the list $\blds = (\bldh_u)_{u \in V'}$.
\Item{(E3)}
Regard $\blds$ as a word of length
$(1{-}r_0) \Delta_1 n \; (= r_\Int R_2 \Delta_2 n)$ over $F$,
and map it by an encoder of $\code_\Int$
into a codeword $\bldw = (\bldw_v)_{v \in V''}$ of $\code_\Int$.
\Item{(E4)}
Using an encoder $\encoder_2 : F^{R_2 \Delta_2} \rightarrow \code_2$,
map $\bldw$ into a codeword $\bldd$ of $\Code_2$ by
\[
(\bldd)_{\!\scriptscriptstyle E_2(v)} \leftarrow \encoder_2(\bldw_v)
\; ,
\quad
v \in V'' \; .
\]
\Item{Output:}
Word $\bldx = (\bldx_u)_{u \in V'}$ in $(\Phi_1 \times \Phi_2)^n$
whose components are given by the pairs
\[
\bldx_u =
\left(
(\bldc)_{\!\scriptscriptstyle E_1(u)},
(\bldd)_{\!\scriptscriptstyle E_2(u)}
\right)
\; ,
\quad u \in V'  \; .
\]
\end{list}
\end{algorithm}
\caption{Encoder for $\BigCode$.}
\label{fig:encoder}
\end{figure*}

\begin{figure*}
\makebox[0in]{}\hrulefill\makebox[0in]{}
\begin{center}
\setlength{\unitlength}{1.25pt}
\begin{picture}(375,220)(000,-15)
%
%
\newsavebox{\pattern}
\newcommand{\drawpattern}{
	\multiput(000,160)(0,-20){5}{\usebox{\pattern}}
	\multiput(000,040)(0,-20){2}{\usebox{\pattern}}
}
\newcommand{\ellipsis}{
        \put(000,055){\circle*{2}}
        \put(000,060){\circle*{2}}
        \put(000,065){\circle*{2}}
}
\newcommand{\drawgraph}{
        \put(003,040){\oval(8,12)[r]}
        \put(037,040){\oval(8,12)[l]}

        \sbox{\pattern}{\circle*{4}}
        \put(000,000){\drawpattern}
        \put(040,000){\drawpattern}

        \put(020,000){\ellipsis}

        \put(000,160){\line(1,-1){40}}
        \put(000,160){\line(2,-1){40}}
        \put(000,160){\line(2,-3){40}}

        \put(000,140){\line(1,-1){40}}
        \put(000,140){\line(2,1){40}}
        \put(000,140){\line(2,-1){40}}

        \put(000,120){\line(1,-1){40}}
        \put(000,120){\line(1,1){40}}
        \put(000,120){\line(2,1){40}}

        \put(000,100){\line(2,-1){40}}
        \put(000,100){\line(1,1){40}}
        \put(000,100){\line(2,1){40}}

        \put(000,080){\line(1,0){40}}
        \put(000,080){\line(2,1){40}}
        \put(000,080){\line(1,2){40}}

        \put(000,080){\line(2,-1){10}}
        \put(040,080){\line(-2,-1){10}}

        \put(000,040){\line(2,1){10}}
        \put(000,040){\line(1,0){12}}
        \put(040,040){\line(-2,1){10}}
        \put(040,040){\line(-1,0){12}}
        \put(000,040){\line(2,-1){40}}

        \put(000,020){\line(1,1){10}}
        \put(040,020){\line(-1,1){10}}
        \put(000,020){\line(2,1){40}}
        \put(000,020){\line(1,0){40}}
}
\newcommand{\ta}{\count230}
\newcommand{\tb}{\count231}
\newcommand{\tc}{\count232}
\newcommand{\ts}{\count233}
\newcommand{\defineword}[1]{
        \ts = 2
        \ta = #1
        \tb = \ta
        \advance \tb by -1
        \tc = \ta
        \multiply \tc by 2
        \put(-2,-\ta){
                \multiput(000,\ts)(0,\ts){\tb}{\line(1,0){4}}
                {\thicklines
                        \framebox(4,\tc){}
                }
        }
}
\put(070,200){\makebox(0,0){\textbf{Step~(E1)}}}
\put(020,205){\line(0,-1){10}}
\sbox{\pattern}{\defineword{6}}
\put(020,000){
        \drawpattern
        \put(000,053){\makebox(0,0){$\bldeta_v$}}
}

\sbox{\pattern}{
        \put(004,000){\vector(1,0){10}}
        \put(014,-06){\framebox(12,12){$\encoder_1$}}
        \put(026,000){\vector(1,0){10}}
}
\put(020,000){\drawpattern}
\put(040,000){\ellipsis}

\sbox{\pattern}{\defineword{9}}
\put(060,000){
        \drawpattern
        \put(000,055){
                \makebox(0,0){$\encoder_1(\bldeta_v)$}}
}

\put(070,000){\drawgraph}
\put(090,180){\makebox(0,0){Graph $\graph_1$}}

\sbox{\pattern}{\defineword{9}}
\put(120,000){
        \drawpattern
        \put(-04,055){
                \makebox(0,0){$(\bldc)_{\scriptscriptstyle E_1(u)}$}}
}
\put(145,200){\makebox(0,0){\textbf{Step~(E2)}}}
\put(120,205){\line(0,-1){10}}
\sbox{\pattern}{
        \put(004,000){\vector(1,0){13}}
        \put(025,000){\circle{16}}
        \put(025,000){\makebox(0,0){$H_0$}}
        \put(033,000){\vector(1,0){13}}
}
\put(120,000){\drawpattern}
\put(145,000){\ellipsis}

\sbox{\pattern}{\defineword{3}}
\put(170,000){
        \drawpattern
        \put(000,050){\makebox(0,0){$\bldh_u$}}
}
\put(215,200){\makebox(0,0){\textbf{Step~(E3)}}}
\put(170,205){\line(0,-1){10}}
\put(174,010){
        \put(000,155){\oval(10,10)[tr]}
        \put(005,085){\line(0,1){70}}
        \put(010,085){\oval(10,10)[bl]}
        \put(010,075){\oval(10,10)[tl]}
        \put(005,075){\line(0,-1){70}}
        \put(000,005){\oval(10,10)[br]}
}
\put(184,090){\vector(1,0){11}}
\put(187,095){\makebox(0,0)[b]{$\blds$}}
\put(195,070){\framebox(40,40){\begin{tabular}{c}
                        Encoder \\ of $\code_\Int$ \end{tabular}}
}
\put(235,090){\vector(1,0){11}}
\put(243,095){\makebox(0,0)[b]{$\bldw$}}

\put(256,010){
        \put(000,155){\oval(10,10)[tl]}
        \put(-05,085){\line(0,1){70}}
        \put(-10,085){\oval(10,10)[br]}
        \put(-10,075){\oval(10,10)[tr]}
        \put(-05,075){\line(0,-1){70}}
        \put(000,005){\oval(10,10)[bl]}
}
\put(310,200){\makebox(0,0){\textbf{Step~(E4)}}}
\put(260,205){\line(0,-1){10}}
\sbox{\pattern}{\defineword{4}}
\put(260,000){
        \drawpattern
        \put(000,050){\makebox(0,0){$\bldw_v$}}
}

\sbox{\pattern}{
        \put(004,000){\vector(1,0){10}}
        \put(014,-06){\framebox(12,12){$\encoder_2$}}
        \put(026,000){\vector(1,0){10}}
}
\put(260,000){\drawpattern}
\put(280,000){\ellipsis}

\sbox{\pattern}{\defineword{6}}
\put(300,000){
        \drawpattern
        \put(000,053){
                \makebox(0,0){$\encoder_2(\bldw_v)$}}
}

\put(310,000){\drawgraph}
\put(330,180){\makebox(0,0){Graph $\graph_2$}}

\sbox{\pattern}{\defineword{6}}
\put(360,000){
        \drawpattern
        \put(-04,053){
                \makebox(0,0){$(\bldd)_{\scriptscriptstyle E_2(u)}$}}
}
\put(360,205){\line(0,-1){10}}
\put(020,-10){\makebox(0,0){\begin{tabular}{c}
                           Information \\ word $\bldeta$ \end{tabular}}}
\put(120,000){\line(0,-1){5}}
\put(125,-05){\oval(10,10)[bl]}
\put(125,-10){\vector(1,0){085}}
\put(240,-10){\makebox(0,0){Codeword $\bldx$}}
\put(360,000){\line(0,-1){5}}
\put(355,-05){\oval(10,10)[br]}
\put(355,-10){\vector(-1,0){085}}
\end{picture}
\end{center}
\makebox[0in]{}\hrulefill\makebox[0in]{}
\caption{Schematic diagram of the encoder for $\BigCode$.}
\label{fig:schematicdiagram}
\end{figure*}

   From the discussion in Section~\ref{sec:outline} and from
the assumption on the code $\code_\Int$ it readily
follows that the encoder in Figure~\ref{fig:encoder}
can be implemented in linear time, where the encoding complexity
per symbol (when measured in operations in $F$)
is $\Poly(\Delta_1,\Delta_2)$. The rate of $\BigCode$
is also easy to compute: the encoder in Figure~\ref{fig:encoder} maps,
in a one-to-one manner,
an information word of length $n$ over an alphabet of size
$q^{R_1 \Delta_1}$,
into a codeword of length $n$ over an alphabet $\Phi$ of size
$q^{\Delta_1 + \Delta_2}$. Thus, the rate of $\BigCode$ is
\begin{equation}
\label{eq:rate2}
\frac{R_1 \Delta_1 n}{(\Delta_1 + \Delta_2) n} =
\frac{R_1}{1 + (\Delta_2/\Delta_1)} \; .
\end{equation}

In the next section, we show
how the parameters of $\BigCode$ can be selected so that
it becomes nearly-MDS and also linear-time decodable.

\subsection{Design, decoding, and analysis}

We will select the parameters of $\BigCode$ quite similarly
to Example~\ref{ex:epsilon^3}.
We assume that the rates $R_1$ and $R_2$ of $\code_1$ and $\code_2$ are 
the same and are equal to some prescribed value $R$, and define
\[
\alpha_R = 8 \cdot (1{-}R) \cdot \max \{ R/\mu, 2/\kappa \}
\]
(notice that $\alpha_R$ can be bounded from above by
a universal constant that does not depend on $R$, e.g.,
by $16/\min \{ 2 \mu, \kappa \}$).
We set $\theta_0 = \kappa \cdot \epsilon$ for some positive
$\epsilon < R$ (in which case $1{-}r_0 < \kappa \cdot \epsilon$),
and then select $q$, $\Delta_1$, and $\Delta_2$ so that
$q > \Delta_1 \ge \alpha_R/\epsilon^3$ and
\begin{equation}
\label{eq:Delta2}
\Delta_2 = \frac{(1{-}r_0) \Delta_1}{r_\Int R} \quad (< \Delta_1) \; ;
\end{equation}
yet we also assume that $q$ is (no larger than) $O(1/\epsilon^3)$.
The graphs $\graph_1$ and $\graph_2$ are taken as Ramanujan graphs
and $\code_0$, $\code_1$, and $\code_2$ are taken as GRS codes over $F$.
(Requiring that both $\Delta_1$ and $\Delta_2$ be valid
degrees of Ramanujan graphs imposes some restrictions
on the value $(1{-}r_0)/(r_\Int R)$. These restrictions
can be satisfied by tuning the precise rate of $\code_\Int$ last.)

Given this choice of parameters, we obtain from~(\ref{eq:Delta2})
that $\Delta_2/\Delta_1 < \epsilon/R$ and, so,
the rate~(\ref{eq:rate2}) of $\BigCode$ is greater than
\begin{equation}
\label{eq:rate_bnd2}
\frac{R}{1 + (\epsilon/R)} > R - \epsilon \; .
\end{equation}
The alphabet size of $\BigCode$ is
\[
|\Phi| = |\Phi_1| \cdot |\Phi_2| = q^{\Delta_1 + \Delta_2} =
 2^{O\left((\log(1/\epsilon))/\epsilon^3 \right)} \; ,
\]
as in~(\ref{eq:alphabetsize}), where we have absorbed
into the $O(\cdot)$ term the constants $\kappa$ and $\mu$.

Our next step in the analysis of the code $\BigCode$ consists
of showing that there exists a linear-time decoder which recovers
correctly any pattern of $t$ errors and $\rho$ erasures, provided that
\begin{equation}
\label{eq:decode0}
2 t + \rho \le (1{-}R{-}\epsilon) n \; .
\end{equation}
This, in turn, will also imply that the relative minimum distance
of $\BigCode$ is greater than $1{-}R{-}\epsilon$,
thus establishing with~(\ref{eq:rate_bnd2}) the fact that $\BigCode$
is nearly-MDS.

Let $\bldx = (\bldx_u)_{u \in V'}$ be the transmitted codeword
of $\BigCode$, where
\[
\bldx_u =
\left(
(\bldc)_{\!\scriptscriptstyle E_1(u)},
(\bldd)_{\!\scriptscriptstyle E_2(u)}
\right)
\; ,
\]
and let $\bldy = (\bldy_u)_{u \in V'}$ be the received word;
each entry $\bldy_u$ takes the form
$(\bldy_{u,1}, \bldy_{u,2})$, where
$\bldy_{u,1} \in \Phi_1 \cup \{ \erasure \}$ and
$\bldy_{u,2} \in \Phi_2 \cup \{ \erasure \}$.
Consider the application of
the algorithm in Figure~\ref{fig:decoder2} to $\bldy$, assuming
that $\bldy$ contains $t$ errors and $\rho$ erasures,
where $2 t + \rho \le (1{-}R{-}\epsilon) n$.

\begin{figure*}
\begin{algorithm}
\settowidth{\Initlabel}{\textbf{Output:}}
\newcommand{\Item}[1]{\item[{\makebox[\Initlabel][c]{\textbf{#1}}}]}
\begin{list}{}{\setlength{\labelwidth}{\Initlabel}%
               \setlength{\leftmargin}{\Initlabel}%
               \addtolength{\itemsep}{1ex}}
\Item{Input:}
Received word $\bldy = (\bldy_u)_{u \in V'}$
in $( \Phi \cup \{ \erasure \} )^n$.
\Item{(D1)}
For $u \in V'$ do:
$\quad \displaystyle (\bldz)_{\!\scriptscriptstyle E_2(u)} \leftarrow
\left\{
\begin{array}{lcl}
\bldy_{u,2}                    && \textrm{if $\bldy_{u,2} \in \Phi_1$}\\
\erasure\erasure\ldots\erasure && \textrm{if $\bldy_{u,2} = \erasure$}
\end{array}
\right. \;$.
\Item{(D2)}
For $v \in V''$ do:
$\quad \displaystyle \tilde{\bldw}_v \leftarrow
\encoder_2^{-1} \left(
    \decoder_2 \left( (\bldz)_{\!\scriptscriptstyle E_2(v)} \right)
\right)$.
\Item{(D3)}
Apply a decoder of $\code_\Int$ to
$\tilde{\bldw} = (\tilde{\bldw}_v)_{v \in V''}$
to produce an information word $\hat{\blds} \in F^{(1-r_0) \Delta_1 n}$.
\Item{(D4)}
Apply a decoder for $(\Code_1(\hat{\blds}))_{\Phi_1}$
to $(\bldy_{u,1})_{u \in V'}$,
as described in Section~\ref{sec:outline},
to produce an information word
$\hat{\bldeta} = (\hat{\bldeta}_v)_{v \in V''}$.
\Item{Output:}
Information word $\hat{\bldeta} = (\hat{\bldeta}_v)_{v \in V''}$
of length $n$ over $F^{R \Delta_1}$.
\end{list}
\end{algorithm}
\caption{Decoder for $(\Code)_\Phi$.}
\label{fig:decoder2}
\end{figure*}

Step~(D1) is the counterpart of the initialization step
in Figure~\ref{fig:decoder} (the entries of $\bldz$ here
are indexed by the edges of $\graph_2$).

The role of Step~(D2) is to compute
a word $\tilde{\bldw} \in \Phi_\Int^n$
that is close to the codeword $\bldw$ of $\code_\Int$,
which was generated in Step~(E3) of Figure~\ref{fig:encoder}.
Step~(D2) uses the inverse of the encoder $\encoder_2$
(which was used in Step~(E4)) and also
a combined error-erasure decoder 
$\decoder_2 : (F \cup \{ \erasure \})^{\Delta_2} \rightarrow \code_2$
that recovers correctly any pattern of $a$ errors (over $F$)
and $b$ erasures, provided that $2a + b < \delta_2 \Delta_2$.
The next lemma provides
an upper bound on the Hamming distance between $\bldw$
and $\tilde{\bldw}$ (as words of length $n$ over $\Phi_\Int$).

\begin{lemma}
\label{lemma:decode}
Under the assumption~(\ref{eq:decode0}),
the Hamming distance between $\bldw$ and $\tilde{\bldw}$
(as words over $\Phi_\Int$) is less than $\mu n$.
\end{lemma}

\emph{Proof:}
Define the function
$\chi : (V' \cup V'') \rightarrow \{ 0, \half, 1 \}$ by
\[
\chi(u) =
\left\{
\begin{array}{lcl} 
1     && \textrm{if $u \in V'$  and $\bldy_{u,2}$ is in error}     \\
\half && \textrm{if $u \in V'$  and $\bldy_{u,2}$ is an erasure}   \\
1     && \textrm{if $u \in V''$ and $\tilde{\bldw}_u \ne \bldw_u$} \\
0     && \textrm{otherwise}
\end{array}
\right. \; .
\]
Assuming that $\tilde{\bldw} \ne \bldw$,
this function satisfies the conditions of Lemma~\ref{lemma:sqrt_st}
with respect to the graph $\graph_2$, where
$\sigma n$ equals $t + (\rho/2)$
and $\tau n$ equals the number of vertices $v \in V''$ such that
$\tilde{\bldw}_v \ne \bldw_v$. By that lemma we get
\begin{eqnarray}
\sqrt{\frac{\sigma}{\tau}} & \ge &
\frac{(\delta_2/2)    - (1{-}\gamma_2) \sigma}{\gamma_2} \ge
\frac{(\delta_2/2)    -  \sigma}{\gamma_2} \nonumber \\
\label{eq:decode}
& > &
\frac{1{-}R - 2 \sigma}{2 \gamma_2} \ge \frac{\epsilon}{2 \gamma_2}
\; ,
\end{eqnarray}
where $\gamma_2$ stands for $\gamma_{\graph_2}$
and the last inequality follows from~(\ref{eq:decode0}).
Now, by~(\ref{eq:Delta2}) we have
\[
\Delta_2 = \frac{(1{-}r_0) \Delta_1}{r_\Int R} >
\frac{\epsilon \Delta_1}{R} \ge
\frac{\alpha_R}{R \cdot \epsilon^2} \ge
\frac{8(1{-}R)}{\mu \cdot \epsilon^2} \; ,
\]
  from which we get the following upper bound on
the square of $\gamma_2$:
\[
\gamma_2^2 \le \frac{4(\Delta_2{-}1)}{\Delta_2^2} <
\frac{4}{\Delta_2} \le
\frac{\mu \cdot \epsilon^2}{2(1{-}R)}\; .
\]
Combining this bound with~(\ref{eq:decode}) yields
\[
\frac{\sigma}{\tau} > \frac{1{-}R}{2 \mu} \; ,
\]
namely, $\tau < 2 \mu \sigma/(1{-}R) < \mu$.\qed

It follows from Lemma~\ref{lemma:decode} that Step~(D2) reduces
the number of errors in $\tilde{\bldw}$ to the extent that allows
a linear-time decoder of $\code_\Int$ to fully recover
the errors in $\tilde{\bldw}$ in Step~(D3).
Hence, the list $\hat{\blds}$, which is computed in Step~(D3),
is identical with the list $\blds$ that was originally encoded in
Step~(E2).

Finally, to show that Step~(D4) yields complete recovery from errors,
we apply Theorem~\ref{thm:main} to the parameters
of the code $(\graph_1,\code_0:\code_1)$.
Here $\theta_0 = \kappa \cdot \epsilon$ and
\[
\gamma_1 = \gamma_{\graph_1} < \frac{2}{\sqrt{\Delta_1}} \le
\frac{2 \epsilon^{3/2}}{\sqrt{\alpha_R}} \le
\frac{\epsilon^{3/2}}{2 \sqrt{(1{-}R)/\kappa}} \; ;
\]
therefore,
\[
\beta = \frac{(\delta_1/2) -
        \gamma_1 \sqrt{\delta_1/\theta_0}}{1 - \gamma_1} >
        \frac{1{-}R}{2} - \gamma_1 \sqrt{\frac{1{-}R}{\theta_0}} >
        \frac{1{-}R{-}\epsilon}{2} \;
\]
and, so, by~(\ref{eq:decode0}), the conditions of
Theorem~\ref{thm:main} hold for $\sigma = (1{-}R{-}\epsilon)/2$
(note that $\beta > 0$ yields $\sqrt{\theta_0 \delta_1}> 2 \gamma_1$,
thus~(\ref{eq:exponentbase}) holds). 

\providecommand{\mynumbering}[1]{\arabic{#1}}
\renewcommand{\mynumbering}[1]{\Alph{#1}}
\useRomanappendicesfalse
\appendices

\section{}
\label{app:A}

We provide here the proofs of
Lemmas~\ref{lemma:alonchung} and~\ref{lemma:sqrt_st}.

Given a bipartite graph $\graph = (V':V'',E)$, we associate
with $\graph$ a $|V'| \times |V''|$ real matrix $X_\graph$
whose rows and columns are indexed
by $V'$ and $V''$, respectively,
and $(X_\graph)_{u,v} = 1$ if and only if 
$\{ u, v \} \in E$.
With a proper ordering on $V' \cup V''$,
the matrix $X_\graph$ is related to
the adjacency matrix of $\graph$ by
\begin{equation}
\label{eq:AX}
A_\graph =
\left(
{\renewcommand{\arraystretch}{1.5}
\begin{array}{c|c}
0 & X_\graph \\
\hline
X_\graph^T & 0 
\end{array} }
\right) \; .
\end{equation}

\begin{lemma}
\label{lemma:eigenvalues_XX}
Let $\graph = (V':V'',E)$ be a bipartite $\Delta$-regular graph
where $|V'| > 1$.
Then $\Delta^2$ is the largest eigenvalue of
the (symmetric) matrix $X_\graph^T X_\graph$ and
the all-one vector $\bldone$ is a corresponding eigenvector.
The second largest eigenvalue of $X_\graph^T X_\graph$ is
$\gamma_\graph^2 \Delta^2$.
\end{lemma}

\emph{Proof:}
We compute the square of $A_\graph$,
\[
A_\graph^2 =
\left(
{\renewcommand{\arraystretch}{1.5}
\begin{array}{c|c}
X_\graph X_\graph^T & 0 \\
\hline
0 & X_\graph^T X_\graph 
\end{array} }
\right) \; ,
\]
and recall the following two known facts:
\begin{list}{}{\settowidth{\labelwidth}{(ii)}}
\item[(i)]
$X_\graph X_\graph^T$ and $X_\graph^T X_\graph$ have the same set of
eigenvalues, each with
the same multiplicity~\cite[Theorem~16.2]{MacDuffee}.
\item[(ii)]
If $\lambda$ is an eigenvalue of $A_\graph$, then so is $-\lambda$,
with the same multiplicity~\cite[Proposition~1.1.4]{DSV}.
\end{list}
We conclude that $\lambda$ is an eigenvalue of $A_\graph$ if and only if
$\lambda^2$ is an eigenvalue $X_\graph^T X_\graph$;
furthermore, when $\lambda \ne 0$,
both these eigenvalues have the same multiplicities in
their respective matrices.
The result readily follows.\qed

For real column vectors $\bldx, \bldy \in \Real^m$,
let $\langle \bldx, \bldy \rangle$ be
the scalar product $\bldx^T \bldy$ and $\| \bldx \|$ be
the norm $\sqrt{\langle \bldx, \bldx \rangle}$.

\begin{lemma}
\label{lemma:1}
Let $\graph = (V':V'',E)$ be a bipartite $\Delta$-regular graph
where $|V'| = n > 1$ and let
$\blds = (s_u)_{u \in V'}$ and
$\bldt = (t_u)_{u \in V''}$ be two column vectors in $\Real^n$.
Denote by $\sigma$ and $\tau$ the averages
\[
\sigma = \frac{1}{n} \sum_{u \in V'}  s_u
\qquad \textrm{and} \qquad 
\tau   = \frac{1}{n} \sum_{u \in V''} t_u \; ,
\]
and let the column vectors $\bldy$ and $\bldz$ in $\Real^n$
be given by
\[
\bldy = \blds - \sigma \cdot \bldone
\qquad \textrm{and} \qquad 
\bldz = \bldt - \tau \cdot \bldone \; .
\]
Define the vector $\bldx \in \Real^{2n}$ by
\[
\bldx = \left( \begin{array}{c} \blds \\ \bldt \end{array} \right) \; . 
\]
Then, 
\[
\left|
\langle \bldx, A_\graph \bldx \rangle - 2 \sigma \tau \Delta n
\right| 
\le 2 \gamma_\graph \Delta \| \bldy \| \cdot \| \bldz \| \; .
\]\vspace{0ex}
\end{lemma}

\emph{Proof:}
First, it is easy to see that
$X_\graph \bldone = X_\graph^T \bldone = \Delta \cdot \bldone$ and that
$\langle \bldy, \bldone \rangle = \langle \bldz, \bldone \rangle = 0$;
these equalities, in turn, yield the relationship:
\[
\langle \bldy, X_\graph \bldz \rangle =
\langle \blds, X_\graph \bldt \rangle - \sigma \tau \Delta n \; .
\]
Secondly, from~(\ref{eq:AX}) we get that
\[
\langle \bldx, A_\graph \bldx \rangle =
2 \langle \blds, X_\graph \bldt \rangle \; .
\]
Hence, the lemma will be proved once we show that
\begin{equation} 
\label{eq:in-prod-1}
\left| \langle \bldy, X_\graph \bldz \rangle \right| \le
\gamma_\graph \Delta \| \bldy \| \cdot \| \bldz \| \; .
\end{equation}

Let
\[
\lambda_1 \ge \lambda_2 \ge \ldots \ge \lambda_n
\]
be the eigenvalues of $X_\graph^T X_\graph$ and let
$\bldv_1, \bldv_2, \ldots, \bldv_n$ be corresponding
orthonormal eigenvectors where, by Lemma~\ref{lemma:eigenvalues_XX},
\[
\lambda_1 = \Delta^2 \; ,
\qquad
\lambda_2 = \gamma_\graph^2 \Delta^2 \; ,
\qquad \textrm{and} \qquad
\bldv_1 = (1/\sqrt{n}) \cdot \bldone \; .
\]
Write 
\[
\bldz = \sum_{i=1}^n \beta_i \bldv_i \; ,
\]
where $\beta_i = \langle \bldz, \bldv_i \rangle$.
Recall, however, that
$\beta_1 = (1/\sqrt{n}) \cdot \langle \bldz, \bldone \rangle = 0$;
so,
\begin{eqnarray*}
\| X_\graph \bldz \|^2 & = &
\langle \bldz, X_\graph^T X_\graph \bldz \rangle \\
& = &
\Bigl\langle
\sum_{i=2}^n \beta_i \bldv_i,
\sum_{i=2}^n \lambda_i \beta_i \bldv_i \Bigr\rangle = 
\sum_{i=2}^n \lambda_i \beta_i^2 \| \bldv_i \|^2 \\
& \le &
\lambda_2 \sum_{i=2}^n \beta_i^2 = \lambda_2 \| \bldz \|^2 =
\gamma_\graph^2 \Delta^2 \| \bldz \|^2 \; .
\end{eqnarray*}
The desired result~(\ref{eq:in-prod-1}) is now obtained from
the Cauchy-Schwartz inequality.\qed

\begin{lemma}
\label{lemma:2}
Let $\graph=(V':V'',E)$ be a bipartite $\Delta$-regular graph
where $|V'| = n > 1$
and let $\chi : (V' \cup V'') \rightarrow \Real$ be
a function on the vertices of $\graph$. Define
the function $w: E \rightarrow \Real$ and
the average $\Eg\{ w \}$ by
\[
w(e) =
\chi(u) \chi(v)
\quad \textrm{for every edge $e = \{ u, v \}$ in $\graph$}
\]
and
\[
\Eg\{ w \} = \frac{1}{\Delta n} \sum_{e \in E} w(e) \; .
\]
Then 
\[
\Big| \Eg \{ w \} - \EAg \{ \chi \} \cdot \EBg \{ \chi \} \Big| \le
\gamma_\graph \sqrt{ \VarAg \{  \chi \} \cdot \VarBg \{  \chi \} } \; ,
\]
where 
\[
\EAg \{ \chi^i \} = \frac{1}{n} \sum_{u \in V'} (\chi(u)^i) \; ,
\]
\[
\EBg \{ \chi^i \} = \frac{1}{n} \sum_{u \in V''} (\chi(u)^i) \; ,
\]
\[
\VarAg \{ \chi \} = \EAg \{ \chi^2 \} - (\EAg \{ \chi \})^2 \; ,
\]
and
\[
\VarBg \{ \chi \} = \EBg \{ \chi^2 \} - (\EBg \{ \chi \})^2 \; .
\]\vspace{0ex}
\end{lemma}

\emph{Proof:} 
Define the column vectors
\[
\blds = (\chi(u))_{u \in V'} \; ,
\quad
\bldt = (\chi(u))_{u \in V''} \; ,
\]
and
\[
\bldx = \left( \begin{array}{c} \blds \\ \bldt \end{array} \right) \; ,
\]
and denote by $\sigma$ and $\tau$ the averages
\[
\sigma = \frac{1}{n} \sum_{u \in V'}  s_u
\qquad \textrm{and} \qquad 
\tau   = \frac{1}{n} \sum_{u \in V''} t_u \; .
\]
The following equalities are easily verified:
\[
\Eg \{ w \} =
\frac{ \langle \bldx, A_\graph \bldx \rangle }{2 \Delta n} \; ,
\]
\[
\EAg \{ \chi \} = \sigma \; ,
\qquad
\EBg \{ \chi \} = \tau \; ,
\]
\[
\VarAg\{ \chi \} =
\frac{1}{n} \cdot \| \blds - \sigma \cdot \bldone \|^2 \; ,
\]
and
\[
\VarBg\{ \chi \} =
\frac{1}{n} \cdot \| \bldt - \tau \cdot \bldone \|^2 \; .
\]
The result now follows from Lemma~\ref{lemma:1}.\qed

\emph{Proof of Lemma~\ref{lemma:alonchung}:}
Using the notation of Lemma~\ref{lemma:2}, write
\begin{equation}
\label{eq:Eg}
\Eg \{ w \} =
\frac{1}{\Delta n}
\sum_{u \in V'} \sum_{v \in \ngbr(u)} \chi(u) \chi(v) \; ,
\end{equation}
\begin{equation}
\EAg \{ \chi \} =
\frac{1}{n} \sum_{u \in V'}  \chi(u) = \sigma \; ,
\end{equation}
and
\begin{equation}
\EBg \{ \chi \} = \frac{1}{n} \sum_{u \in V''} \chi(u) = \tau \; .
\end{equation}
Since the range of $\chi$ is restricted to the interval $[0,1]$, we have
\[
\EAg \{ \chi^2 \} \le \EAg \{ \chi \}
\qquad \textrm{and} \qquad
\EBg \{ \chi^2 \} \le \EBg \{ \chi \} \; ;
\]
hence, the values $\VarAg \{ \chi \}$ and $\VarBg \{ \chi \}$
can be bounded from above by
\begin{equation}
\label{eq:VarAB}
\VarAg \{ \chi \} \le \sigma - \sigma^2
\qquad \textrm{and} \qquad
\VarBg \{ \chi \} \le \tau - \tau^2 \; .
\end{equation}
Substituting~(\ref{eq:Eg})--(\ref{eq:VarAB}) into
Lemma~\ref{lemma:2} yields
\[
\Biggl| \frac{1}{\Delta n}
\Bigl( \sum_{u \in V'} \sum_{v \in \ngbr(u)}
\chi(u) \chi(v) \Bigr) - \sigma \tau \Biggr| \le 
\gamma_\graph \sqrt{\sigma (1{-}\sigma) \tau (1{-}\tau)} \; ;
\]
so,
\begin{eqnarray*}
\lefteqn{
\frac{1}{\Delta n}
\sum_{u \in V'} \sum_{v \in \ngbr(u)} \chi(u) \chi(v) }
\makebox[5ex]{} \\
& \le & 
\sigma \tau +
\gamma_\graph \sqrt{\sigma (1{-}\sigma) \tau (1{-}\tau)} \\
& = &
(1{-}\gamma_\graph) \sigma \tau +
\gamma_\graph \sqrt{\sigma \tau} \left( \sqrt{\sigma \tau} 
+ \sqrt{(1{-}\sigma) (1{-}\tau)} \right) \\
& \le &
(1{-}\gamma_\graph) \sigma \tau + \gamma_\graph \sqrt{\sigma \tau}
\; ,
\end{eqnarray*}
as claimed.\qed
 
\emph{Proof of Lemma~\ref{lemma:sqrt_st}:}
We compute lower and upper bounds on the average
\[
\frac{1}{\Delta n}
\sum_{v \in V''} \sum_{u \in \ngbr(v)} \chi(u) \chi(v) \; .
\]
On the one hand, this average equals
\[
\frac{1}{\Delta n}
\sum_{\scriptstyle v \in V'' : \atop \scriptstyle \chi(v) > 0} \chi(v)
\underbrace{\sum_{u \in \ngbr(v)} \chi(u)}_{{} \ge \delta \Delta/2} \ge
\frac{1}{\Delta n} \cdot \frac{\delta \Delta}{2} 
\underbrace{\sum_{\scriptstyle v \in V''} \chi(v)}_{\tau n} =
\frac{\delta \tau}{2} \; ,
\]
where the inequality follows from the assumed conditions on $\chi$.
On the other hand, this average also equals
\[
\frac{1}{\Delta n}
\sum_{u \in V'} \sum_{v \in \ngbr(u)} \chi(u) \chi(v) \le
(1{-}\gamma_\graph) \sigma \tau +
\gamma_\graph \sqrt{\sigma \tau} \; ,
\]
where the inequality follows from Lemma~\ref{lemma:alonchung}.
Combining these two bounds we get
\[
\frac{\delta \tau}{2} \le 
(1{-}\gamma_\graph) \sigma \tau +
\gamma_\graph \sqrt{\sigma \tau} \; , 
\]
and the result is now obtained by dividing by $\gamma_\graph \tau$
and re-arranging terms.\qed

\section{}
\label{app:B}

When analyzing the complexity of the algorithm in Figure~\ref{fig:decoder}, 
one can notice that the decoder $\decoder \in \{ \decoder', \decoder'' \}$
needs to be applied at vertex $u$, only if $(\bldz)_{E(u)}$ has been 
modified since the last application of $\decoder$ at that vertex. Based on 
this observation, we prove the following lemma. 

\begin{lemma} 
The number of (actual) applications of the decoders $\decoder'$ and $\decoder''$
in the algorithm in Figure~\ref{fig:decoder} can be bounded from above by 
$\omega \cdot n$, where 
\[
\omega = 2 \cdot \left\lceil
\frac{\log\left( \frac{\displaystyle
\Delta \beta \sqrt{\sigma}}{ \displaystyle {\beta} - \sigma} \right)}
{\log \left( \displaystyle
\frac{\theta \delta}{4 \gamma_\graph^2} \right) } \right\rceil
+ \frac{\displaystyle 1 + \frac{\theta}{\delta} } 
{\displaystyle 1 -
\left( \frac{4 \gamma_\graph^2}{\theta \delta} \right)^2 } \; .
\]
\label{lemma:complexity}
\end{lemma} 

\emph{Proof:}
Define $i_T$ by 
\[
i_T \; = \;
2 \cdot 
\left\lceil \frac{\log\left( \frac{\displaystyle
\Delta \beta \sqrt{\sigma}}{ \displaystyle {\beta} - \sigma} \right)}
{\log \left( \displaystyle \frac{\theta \delta}{4 \gamma_\graph^2}
\right) } \right\rceil
 \; . 
\]
It is easy to verify that 
\begin{equation}
\left( \frac{\theta \delta}{4 \gamma_\graph^2}
\right)^{i_T/2} \left( \frac{1}{\sqrt{\sigma}} -
\frac{\sqrt{\sigma}}{\beta} \right)
\; \ge \; \Delta\; .
\label{eq:cond-i-t}
\end{equation}
In the first $i_T$ iterations in Figure~\ref{fig:decoder}, we apply  
the decoder $\decoder$ (which is either $\decoder'$ or $\decoder''$)
at most $i_T \cdot n$ times. 

Next, we evaluate the total number of applications of the decoder
$\decoder$ in iterations $i = i_T+1, i_T+2, \cdots, \nu$.
We hereafter use the notations $U_i$ and $S_i$ as in the proof of
Theorem~\ref{thm:main}.
Recall that we need to apply the decoder $\decoder$ to 
$(\bldz)_{E(u)}$ for a vertex $u \in U_{i+2}$,
only if at least one entry in $(\bldz)_{E(u)}$
--- say, the one that is indexed by the edge
$\{u, v \} \in E(u)$ --- has been altered during iteration $i+1$.
Such an alteration may occur only 
if $v$ is a vertex in $U_{i+1}$ with an adjacent 
vertex in $S_i$. We conclude that $\decoder$ needs to be applied
at vertex $u$ during iteration $i+2$ 
only if $u \in \ngbr(\ngbr(S_i))$. The number of such vertices $u$,
in turn, is at most $\Delta^2 \, |S_i| = \Delta^2 \cdot \sigma_i n$. 
 
We now sum the values of $\Delta^2 \sigma_i n$ over iterations $i = i_T+1, i_T+2, \cdots, \nu$:
\begin{eqnarray}
\lefteqn{
\Delta^2 n \cdot \sum_{i=i_T+1}^{\nu } \sigma_i } \makebox[5ex]{}
\nonumber \\
& = &
\Delta^2 n
\left( \sum_{j=i_T/2}^{\lfloor (\nu-1)/2 \rfloor} \sigma_{2j+1} 
+ \sum_{j=i_T/2}^{{\lfloor (\nu-2)/2 \rfloor}} \sigma_{2j+2} \right)
\nonumber \\
& \le &
\Delta^2 n \cdot
\sum_{j=i_T/2}^{{\lfloor (\nu-1)/2 \rfloor}} \sigma_{2j+1}
\left( 1 + \frac{\theta}{\delta} \right) \; , 
\label{eq:sum-sigma}
\end{eqnarray}
where the last inequality is due to~(\ref{eq:sigmas-ratio}). 

  From~(\ref{eq:recurrence}) (and by neglecting a positive term), we obtain 
\[
\frac{1}{\sqrt{\sigma_{i+1}}} \; \ge \;
\left( \frac{\theta \delta}{4 \gamma_\graph^2}
\right)^{i/2} \left(\frac{1}{\sqrt{\sigma}} -
\frac{\sqrt{\sigma}}{\beta} \right) \; 
\]
for even $i \ge i_T$.
Therefore, the expression in~(\ref{eq:sum-sigma}) is bounded 
from above by 
\begin{eqnarray*}
\lefteqn{
\frac{\displaystyle \Delta^2 n
\left( 1 + \frac{\theta}{\delta} \right) \cdot  
\left( \frac{4 \gamma_\graph^2}{\theta \delta} \right)^{i_T}}
{\displaystyle \left( 1 -
\left( \frac{4 \gamma_\graph^2}{\theta \delta} \right)^2 \right)
\left(\frac{1}{\sqrt{\sigma}} - \frac{\sqrt{\sigma}}{\beta} \right)^2} 
} \makebox[10ex]{} \nonumber \\
& \le &
\frac{\displaystyle \Delta^2 n
\left( 1 + \frac{\theta}{\delta} \right) \cdot \frac{1}{\Delta^2}} 
{\displaystyle 1 - \left( \frac{4 \gamma_\graph^2}{\theta \delta}
\right)^2 } \nonumber \\
& = &
\frac{\displaystyle n \left( 1 + \frac{\theta}{\delta} \right) } 
{\displaystyle 1 -
\left( \frac{4 \gamma_\graph^2}{\theta \delta} \right)^2 } \; ,
\end{eqnarray*}
where the inequality follows from~(\ref{eq:cond-i-t}). 

Adding now the number of applications of the decoder $\decoder$ during the 
first $i_T$ iterations, we
conclude that the total number of applications of the decoder $\decoder$ is
at most $\omega \cdot n$, where  
\[
\omega = i_T + 
\frac{\displaystyle  1 + \frac{\theta}{\delta} } 
{\displaystyle 1 - \left( \frac{4 \gamma_\graph^2}{\theta \delta}
\right)^2 } \; .
\]
\qed

\end{document}